# Absent daddy, but important father

Nathanaël M.A. Jacquier[1,2], Thomas Widiez[1*]

[1] Laboratoire Reproduction et Développement des Plantes, Univ Lyon, ENS de Lyon, UCB Lyon 1, CNRS, INRAE, F-69342, Lyon, France.
[2] Limagrain, Limagrain Field Seeds, Research Centre, F-63360 Gerzat, France.
Email: thomas.widiez@ens-lyon.fr

**Abstract**:
Mixing maternal and paternal genomes is the base of plant sexual reproduction, but some so-called "haploid inducer lines" lead to the formation of seeds with well-developed embryos with solely the maternal genome. A recent study added a new piece to the puzzle of this enigmatic *in planta* haploid embryo induction process.

Like most eukaryotes, plants carry a dual genome of paternal and maternal origins. Sexual reproduction allows mixing of genetic information, which produces diversity that enables breeding of new plant varieties with improved agronomic traits. However, plant breeding process often requires homozygous or inbred lines with fixed genetic material in order to evaluate the performances of various genetic combinations. To generate these lines in conventional breeding is a time-consuming process which needs multiple-generation of selfing. The production of doubled haploids plants represents a shortcut to obtaining lines of genome homozygosity, which could be achieved in two rather than six or more generations[1]. Maize breeding has benefited greatly from this doubled haploid technology thanks to the haploid inducer lines that induce the formation of haploid embryos in seeds[1] (**Fig. 1a**). The embryos germinate into haploid seedlings carrying a single set of maternal chromosomes[1]. Recently, haploid inducer lines have been also elegantly repurposed as powerful tool to deliver genome-editing machinery into commercial crop varieties that are recalcitrant to transformation techniques[1,2]. Despite being a powerful tool in plant breeding and research applications, the molecular basis behind *in planta* haploid induction remains fragmentary[1]. In this issue of *Nature Plants,* Li et al.[3] found that mutating the phospholipase D3 gene (*ZmPLD3*) can induce maternal haploid embryos, adding a new piece of understanding about this intriguing and useful biological process.

Genetic architecture of maize haploid inducer lines is complex, and up to now only two molecular players have been identified by quantitative trait locus (QTL) mapping. The *MATRILINEAL/NOT LIKE DAD/ZmPHOSPHOLIPASE-A1* (*MTL/NLD/ZmPLA1*) gene was first identified as a key regulator whose mutation causes haploid induction in maize[4–6]. Later on, the mutation in the *DOMAIN MEMBRANE PROTEIN* gene (*ZmDMP*) was shown to increase by up to six-fold the haploid induction rate when combined with the *mtl/nld/zmpla1* mutation[7]. Since *MTL/NLD/ZmPLA1* is a pollen specific gene belonging to the phospholipase family[4–6], Li and colleagues conducted data mining into literature to look for other pollen specifically expressed phospholipases genes with impaired expression in *mtl/nld/zmpla1* mutant background, with the hope to find other regulators for haploid induction[3]. As a result, they identified the *ZmPLD3* gene which showed the expected expression pattern, and experimentally validated that mutation of *ZmPLD3* led to haploid embryo induction with ~1 % of the kernels carrying haploid embryos.

The underlying mode of action of maize haploid induction remains elusive, but the fact that two players are phospholipases suggests the involvement of phospholipid metabolism. Indeed, phospholipases are enzymes that hydrolyze phospholipids, and depending on the phospholipase type they produce different products. As a member of the phospholipase A family, MTL/NLD/ZmPLA1 should generate free fatty acids and lysophospholipids (**Fig. 1**), whereas ZmPLD3, as a phospholipase D class member, should hydrolyze the headgroup of phospholipid and release phosphatidic acid (**Fig. 1**). Considering the diverse roles of phospholipases in both structural membrane organization and lipid cell signaling[8], it is difficult to establish, at this stage, a mechanistic link between the phospholipids and the final phenotype of haploid embryo induction. Thus, more work to identify the exact substrates of these phospholipases will surely help to solve the puzzle. Using protoplast transient system, Li et al.[3] found ZmPLD3 in different cellular compartments, including endoplasmic reticulum, Golgi complex, cytosol and plastid. This relatively broad subcellular localization makes it difficult to define ZmPLD3's specific function in haploid induction, but it can be speculated that impairing its activity in pollen context could lead to drastic change in the pollen phospholipids landscape. Thus, investigations about the temporal and spatial expression of ZmPLD3 within its place of action, the pollen grain, may bring key clues: Does ZmPLD3 acts from the sperm cells or from the neighboring vegetative cell (**Fig. 1b**)? Interestingly, contrary to previous reports[4,5], the MTL/NLD/ZmPLA1 protein was recently demonstrated to not localize inside but outside of the sperm cells, on the poorly characterized endo-plasma membrane that encircles the sperm cells[9] (**Fig. 1b**). A role of ZmPLD3 on this pollen endo-plasma membrane should be thus consider, as this unique membrane might be of critical importance for a proper communication between the pollen vegetative cell and its enclosed sperm cells, and possibly for haploid induction phenotype. More broadly, spatial clarifications of lipid metabolic defects occurring in pollen grains leading to haploid induction are needed to elucidate the roles of those two phospholipases.

To better decipher the genetic interactions between this new phospholipase (*ZmPLD3*) and previously known molecular components (*MTL/NLD/ZmPLA1* and *ZmDMP*), Li and colleagues[3] combined the different mutations. Interestingly, they identified a synergetic interaction on haploid induction rate (HIR) between the mutations of both phospholipases (*MTL/NLD/ZmPLA1* and *ZmPLD3*). The HIR goes from ~1% for each single mutant to ~4% HIR for *mtl/nld/zmpla1-zmpld3* double mutant. Genetic relationship between the phospholipases and ZmDMP seems more complex, since mutation of *zmdmp* increased the HIR (to ~5%) when combined with *mtl/nld/zmpla1*, but not when combined with *zmpld3*. However, combining the three mutations boosted the HIR even higher to ~7%, but *zmpld3* remained at a heterozygous state. Indeed, the authors were not able to obtain *zmpld3* mutation in homozygous state in this genetic background due to a strong segregation biased. This result implies that stacking different mutations impacts pollen fitness or impairs the success of fertilization. The fact that favorable alleles could not be obtained at a homozygous state undoubtedly causes problems for the development of new haploid inducer lines. Knowing the exact mode of action of these molecular players and the detailed mechanisms behind maize haploid induction would certainly facilitate to find a balance between damage of pollen function and haploid induction capacity. It could then be helpful for designing super high haploid inducer lines, and also for translating this important property to other crops, especially dicots.

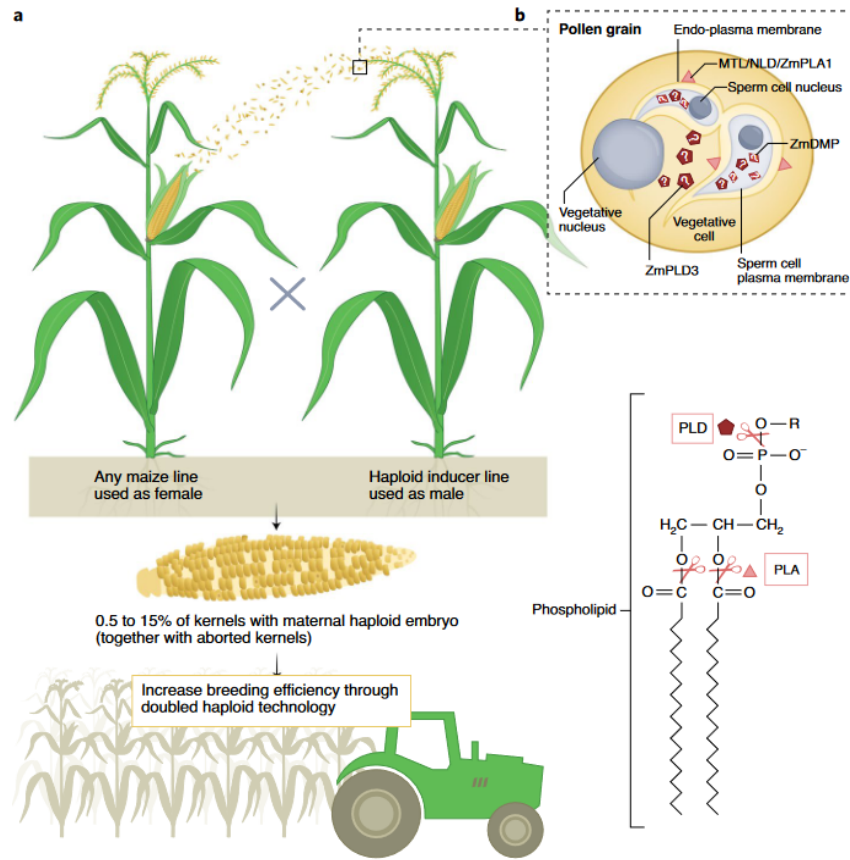

**Fig. 1: Mutation in the phospholipase *ZmPLD3* lead to haploid embryo induction in maize**.

**a**, Maize haploid inducer lines, when used as male parent (pollen), have the ability to trigger the development of kernels having embryo with solely the genetic information from the female parent. This fully *in planta* process represents the first step of doubled haploid technology that increases the plant breeding efficiency. **b**, the three molecular players identified in maize haploid embryo induction process are MATRILINEAL/NOT LIKE DAD/ZmPHOSPHOLIPASE-A1 (MTL/NLD/ZmPLA1), the DOMAIN MEMBRANE PROTEIN (ZmDMP) and the newly identified phospholipase D3 (ZmPLD3). MTL/NLD/ZmPLA1 and ZmPLD3 belong to two different family of phospholipases (PLs) family, enzymes that cleave phospholipids, the major component of cellular membranes.